\documentclass[sigconf,nonacm, natbib=true]{acmart}
\renewcommand\footnotetextcopyrightpermission[1]{}
\usepackage{enumitem}
\usepackage{algorithm}
\usepackage{algpseudocode}
\usepackage{lipsum}  %
\usepackage{afterpage}  %
\usepackage{booktabs}  %
\usepackage{multirow}
\usepackage{graphicx}
\usepackage{stfloats}
\usepackage{diagbox}
\usepackage{dsfont}
\usepackage{subcaption}
\usepackage{cleveref}
\usepackage{array, arydshln} %
\usepackage{balance}
\AtBeginDocument{%
  }

\begin{document}

\title{Response Quality Assessment for Retrieval-Augmented Generation via Conditional Conformal Factuality}

\author{Naihe Feng}
\email{nfeng@dal.ca}
\orcid{0009-0006-8815-7716}
\affiliation{%
  \institution{Dalhousie University}
   \city{Halifax}
  \state{NS}
  \country{Canada}
}
\author{Yi Sui}
\email{amy@layer6.ai}
\orcid{0009-0009-9207-7403}
\affiliation{%
  \institution{Layer 6 AI}
  \city{Toronto}
  \state{ON}
  \country{Canada}
}
\author{Shiyi Hou}
\email{gloria@layer6.ai}
\orcid{0009-0000-8445-8577}
\affiliation{%
  \institution{Layer 6 AI}
  \city{Toronto}
  \state{ON}
  \country{Canada}
}
\author{Jesse C. Cresswell}
\email{jesse@layer6.ai}
\orcid{0000-0002-9284-8804}
\affiliation{%
  \institution{Layer 6 AI}
  \city{Toronto}
  \state{ON}
  \country{Canada}
}
\author{Ga Wu}
\email{ga.wu@dal.ca}
\orcid{0000-0002-0370-0622}
\affiliation{%
  \institution{Dalhousie University}
   \city{Halifax}
  \state{NS}
  \country{Canada}
}

\renewcommand{\shortauthors}{Naihe Feng, Yi Sui, Shiyi Hou, Jesse C. Cresswell, and Ga Wu}

\begin{abstract}

Existing research on Retrieval-Augmented Generation (RAG) primarily focuses on improving overall question-answering accuracy, often overlooking the quality of sub-claims within generated responses. Recent methods that attempt to improve RAG trustworthiness, such as through auto-evaluation metrics, lack probabilistic guarantees or require ground truth answers. To address these limitations, we propose Conformal-RAG, a novel framework inspired by recent applications of conformal prediction (CP) on large language models (LLMs). Conformal-RAG leverages CP and internal information from the RAG mechanism to offer statistical guarantees on response quality. It ensures group-conditional coverage spanning multiple sub-domains without requiring manual labelling of conformal sets, making it suitable for complex RAG applications. Compared to existing RAG auto-evaluation methods, Conformal-RAG offers statistical guarantees on the quality of refined sub-claims, ensuring response reliability without the need for ground truth answers. Additionally, our experiments demonstrate that by leveraging information from the RAG system, Conformal-RAG retains up to 60\% more high-quality sub-claims from the response compared to direct applications of CP to LLMs, while maintaining the same reliability guarantee.\footnote{Pre-print Accepted by SIGIR 2025} 
\end{abstract}


\maketitle
\footnotetext[2]{GitHub: \href{https://github.com/n4feng/ResponseQualityAssessment}{github.com/n4feng/ResponseQualityAssessment}}

\section{Introduction}
Existing research in Retrieval-Augmented Generation (RAG)~\cite{lewis2020rag, gao2024retrievalaugmentedgenerationlargelanguage} mostly focuses on improving overall question-answering accuracy \cite{yu2024evaluationretrievalaugmentedgenerationsurvey}, but often overlooks the quality of sub-claims within generated responses, leading to partially incorrect outputs and hard-to-detect errors \cite{min2023factscorefinegrainedatomicevaluation}. Human evaluations reveal that RAG-based question-answering systems sometimes misinterpret user queries \cite{agrawal2024mindfulragstudypointsfailure, wu-etal-2024-synchronous}, struggle with reasoning in unseen scenarios \cite{mirzadeh2024gsmsymbolicunderstandinglimitationsmathematical, huang-chang-2023-towards}, and may generate claims that are irrelevant or even contradictory to the provided documents~\cite{niu-etal-2024-ragtruth, wu2024llmsciterelevantmedical}.

\begin{figure}[t]
    \centering    \includegraphics[width=\columnwidth]{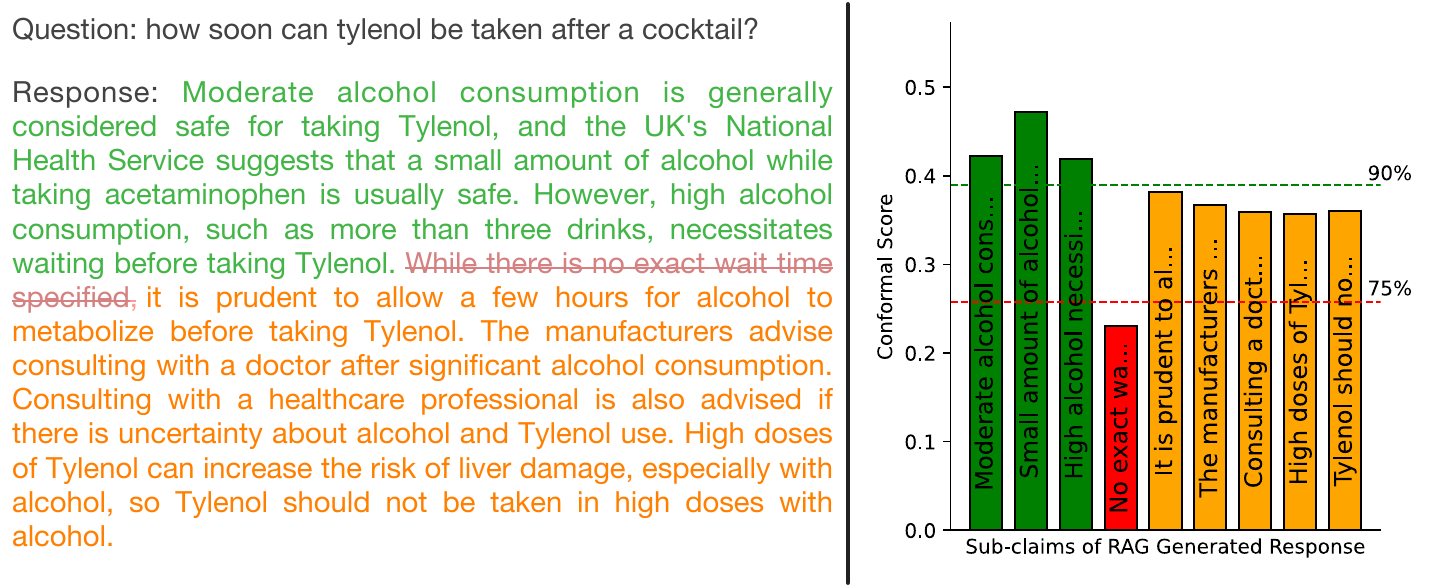}
    \vspace{-6mm}
    \caption{Conformal-RAG filters RAG's responses based on a calibrated factuality threshold. We show two example thresholds guaranteeing 75\% and 90\% factuality. Claims with scores below the threshold are removed from the final response.}
    \label{fig:demo}
    \vspace{-2mm}
\end{figure}

Ensuring the trustworthiness of RAG systems remains a challenge, prompting research into various evaluation solutions.
One straightforward way to quantify the trustworthiness of RAG systems is through auto-evaluation based on well-defined metrics. Unfortunately, popular auto-evaluation methods require ground truth answers at inference time, making them impractical in real applications~\cite{song2024measuringenhancingtrustworthinessllms, ru2024ragcheckerfinegrainedframeworkdiagnosing}. While some research has addressed this problem~\cite{es2023ragasautomatedevaluationretrieval, saadfalcon2024aresautomatedevaluationframework}, auto-evaluation methods still face criticism due to their lack of probabilistic guarantees. 
Compared to the evaluation techniques mentioned above, conformal prediction provides a stronger theoretical foundation for ensuring soundness of evaluations through statistical guarantees. 
In hallucination detection tasks, conformal factuality has provided remarkably robust guarantees on large language model (LLM) outputs, solely relying on the LLM's parametric knowledge~\cite{quach2024conformal, mohri2024languagemodelsconformalfactuality, cherian2024llmenhancedconformal}. Although recent work has integrated conformal prediction into RAG systems~\cite{kang2024cragcertifiedgenerationrisks}, it primarily focuses on analyzing generation risks based on adjustable parameters rather than verifying the factuality of sub-claims, leaving a critical research gap unfilled.

This paper presents Conformal-RAG, a conformal prediction~\cite{vovk2005algorithmic, angelopoulos2022gentleintroductionconformalprediction, cresswell2024} framework tailored for RAG systems. The proposed framework leverages contextual information (retrieved external knowledge) from a RAG system, and a high-quality conformal scoring function, leading to substantially more retained response content compared to existing solutions when targeting the same factuality threshold. In particular, Conformal-RAG can ensure group-conditional factuality \cite{vovk2003mondrian, lei2013distribution, foygelbarber2020limits} spanning multiple sub-domains without requiring manual annotation of conformal set validity, making it highly adaptable for complex RAG applications. We empirically evaluate Conformal-RAG on four benchmark datasets from two domains, Wikipedia \cite{mallen-etal-2023-trust,min2023factscorefinegrainedatomicevaluation,yang2018hotpotqadatasetdiverseexplainable} and medicine~\cite{jeong2024olaph}. The experimental results show that Conformal-RAG retains up to $60\%$ more sub-claims from the output in question-answering tasks for the same factuality level compared to existing baselines. 

\section{Preliminaries and Related Work}
Here, we briefly review conformal prediction and its role in ensuring the trustworthiness of question-answering (QA) tasks. Due to space constraints, we do not cover the broader literature on RAG system trustworthiness, as comprehensive surveys already provide an up-to-date literature review~\cite{zhou2024trustworthiness}.

\subsubsection*{Conformal Prediction}
Conformal Prediction (CP) \cite{vovk2005algorithmic} is a statistical framework that transforms heuristic uncertainty estimates into rigorous, calibrated confidence measures. It provides coverage guarantees over prediction sets, where larger sets indicate higher model uncertainty \cite{angelopoulos2022gentleintroductionconformalprediction}. For a prediction task with possible outputs $Y$, given a conformity measure \( S \) and a tolerable error level \( \alpha \), the conformal prediction set for a new example \( x_\text{test} \) is
\begin{equation}\label{eq:prediction_set}
C_{\hat q}(x_\text{test}) = \{ y\in Y \mid S(x_\text{test}, y) \leq {\hat q} \},
\end{equation}
where \( {\hat q} \) is the $\frac{\lceil(n+1)(1-\alpha)\rceil}{n}$-quantile of scores $S$ over a calibration dataset containing $n$ datapoints. When calibration and test data are drawn i.i.d. from a distribution $\mathbb{P}$, CP guarantees marginal coverage
\begin{equation}\label{eq:coverage}
    \mathbb{P}(y^*_\text{test}\in C_{\hat q}(x_\text{test})) \geq 1-\alpha.
\end{equation}

\subsubsection*{Conformal Factuality for Open-ended QA} 
In classification tasks where $Y$ is a finite label set, CP is straightforward to apply. However, in generative settings like open-ended QA, the output space is effectively infinite, with many semantically equivalent responses. One approach to constrain this space is to limit the output token count \cite{kang2024cragcertifiedgenerationrisks}, however, explicit token limits are not well-suited for open-ended QA, where responses vary in length and structure.

A more principled approach to factuality assessment is to construct prediction sets implicitly as the set of all statements that entail the model's output \cite{mohri2024languagemodelsconformalfactuality}. An output $y$ is factual if the ground truth $y^*$ entails it (denoted by $y^*\Rightarrow y)$, and CP enables calibration of the model's confidence about factuality. Inspired by FActScore~\cite{min2023factscorefinegrainedatomicevaluation}, for long-form answers with multiple claims, one may estimate factuality per claim, filtering out low-confidence ones based on a threshold $\hat q$, while ensuring retained claims meet a factuality guarantee
\begin{equation}
    \mathbb{P}(y^*_\text{test} \Rightarrow  y_\text{test}(x_\text{test}; \hat q)) \geq 1-\alpha.
    \label{eq:conformal_factuality}
\end{equation}
Despite the remarkable probabilistic guarantee offered by conformal factuality, LLMs relying solely on parametric knowledge often generate non-factual statements \cite{maynez2020faithfulness} and struggle with confidence calibration \cite{xiong2024can}, which leads to high claim-rejection rates under strict factuality thresholds. While level-adaptive conformal prediction helps retain more claims, it comes with the cost of reducing overall factuality rates \cite{cherian2024llmenhancedconformal}.

\section{Methodology}
We introduce Conformal-RAG, a framework leveraging CP and the RAG mechanism to offer statistical guarantees on response quality while remaining grounded in documents containing domain knowledge. Below we discuss the end-to-end application of the framework, followed by an in-depth examination of how concepts from CP are applied.

\subsection{Conformal Factuality for RAG}
\label{subsec:e2e_application}

\subsubsection*{Problem Formulation}
Given a query $x\in X$, a RAG model retrieves a set of $m$ relevant documents $D=\{d_1, d_2,...,d_m\}$ from its knowledge corpus. The model then generates an answer $\hat{y}$ composed of $p$ sub-claims $\hat{y}=\{c_1, c_2, ...., c_{p}\}$. The goal of Conformal-RAG is to modify $\hat{y}$ by filtering out sub-claims, producing $y$ which satisfies \cref{eq:conformal_factuality} where $\alpha$ is the predefined error tolerance level, and $y$ consists of a subset of claims from $\hat{y}$, i.e. $y\subseteq \hat{y}=\{c_1, c_2, ...., c_{p}\}$.

\vspace{-2pt}\subsubsection*{Context Similarity-based Conformal Score} 
The first step of our method is to design and calibrate a function to score the relevance of claims. For each query $x$ in the calibration set, we obtain the generated answer from RAG as $\hat{y}=\{c_1, c_2, ...., c_{p}\}$. Our scoring function $R(c\in \hat{y})$ assigns each claim $c$ a relevance score as shown in \cref{alg:rag_score}. First we compute the cosine similarity between the claim and each of the $m$ retrieved documents. These similarity scores are then multiplied by the cosine similarity between the corresponding document and the original query. Finally, the relevance score $R(c)$ takes the maximum of these values across all $m$ documents (or zero if all scores are negative).

\begin{algorithm}[t]
\caption{RAG Sub-claim Scoring}\label{alg:rag_score}
\begin{algorithmic}[1]
\Require Query $x$, retrieved documents $D=\{d_1, d_2,...d_m\}$, generated answer $\hat{y}=\{c_1, c_2, ...., c_{p}\}$.
\For{$c_k \in \hat{y}$}
    \For{$d_j\in D$}
        \State $s_{kj} = \text{CosineSimilarity}(x, d_j) \cdot \text{CosineSimilarity}(c_k, d_j)$
    \EndFor
    \State $r_k = \max(\{s_{kj}\}_{j=1}^m \cup 0)$ \Comment{Sub-claim relevance scores}
\EndFor
\State \Return $\{r_k\}_{k=1}^{p}$
\end{algorithmic}
\end{algorithm}
\subsubsection*{Automatic Calibration Set Annotation} The second step is to design an annotation function which takes advantage of the ground-truth answers from the calibration set to judge the factuality of claims. Specifically, we prompt an LLM~\cite{gu2024survey} to annotate if a given sub-claim is factual by providing the query $x$, ground-truth answer $y^*$, as well as the retrieved documents $D$. The annotation function $A(c\in\hat{y}, x, y^*, D) = 1$ when the sub-claim $c$ is factual and $A(c\in\hat{y}, x, y^*, D) = 0$ when it is non-factual. 

\vspace{-2pt}\subsubsection*{Inference} Based on the relevance scores and annotations generated for each claim across queries in the calibration dataset, we apply CP to calibrate a threshold $\hat q$. Details on the marginal and conditional CP approaches are given below in \cref{subsec:marginal_conformal_rag} and \cref{subsec:conditional_converage_rag}. At inference, only queries and documents are available. Sub-claims and relevance scores are generated in the same way as during calibration. Then, claims are removed from the generated answer if their relevance is below the calibrated threshold,  creating the conformally factual output ${y}(x; \hat q) = \{c\in \hat{y} \mid R(c) \geq \hat q\}$.

Note that LLM-generated answers may not always be in the form of clearly separated sub-claims. Following previous work \cite{mohri2024languagemodelsconformalfactuality}, we use an LLM to decompose the answer into sub-claims. Similarly, since removing sub-claims may affect the grammatical structure of the overall answer, the final set of claims is fed back into an LLM, which is prompted to merge them into a coherent response.

\subsection{Marginal Conformal Factuality with RAG}
\label{subsec:marginal_conformal_rag}
Our marginal CP calibration builds off of work by \citet{mohri2024languagemodelsconformalfactuality}, but takes advantage of the RAG mechanism through our relevance scoring function. Our aim is to guarantee factuality of generated answers in the sense that the final generated output is entailed by the ground truth answer $y^*_\text{test}$ with high probability, satisfying \cref{eq:conformal_factuality}. 

We introduce a filtering function $F_q(\{ c \})$ acting on a set of claims, and satisfying both $F_0(\{ c \}) = \{c\}$ and $F_\infty(\{ c \}) = \emptyset$. As the threshold $q$ increases from 0, $F_q$ progressively filters out more of the claims, and hence satisfies a nesting property: $F_q(\{c\}) \subseteq F_{q'}(\{c\})$ for $q\geq q'$ \cite{Gupta_2022}. The filtering function is constructed using the relevance scores $R(c)$ described in \cref{alg:rag_score} as
\begin{equation}\label{eq:filter}
    F_q(\hat y) = \{c\in \hat y \mid R(c) \geq q\}.
\end{equation}
To determine the appropriate threshold $q$ we use CP calibration over the conformal scores
\begin{equation}\label{eq:conformal_factuality_score}
S(x_i, y_i^*) := \inf \{ q \in \mathbb{R}^+ \mid \forall q'\geq q, \forall c\in F_{q'}(\hat y_i), 
A(c, x_i, y_i^*, D) = 1\}.
\end{equation} 
That is, the score $S$ is the smallest threshold $q$ such that all retained claims are considered factual by the annotation function $A$ from \cref{subsec:e2e_application}.
Then, the conformal threshold $\hat{q}$ is set as the $\frac{\lceil(n+1)(1-\alpha)\rceil}{n}$ quantile of the conformal scores over the calibration set.

On inference data we filter out claims with relevance score $R(c)$ less than $\hat q$, i.e. we return $y_\text{test} = F_{\hat q}(\hat y)$. Under the assumption that the annotation function is correct on the calibration data, these sets of filtered claims will satisfy \cref{eq:conformal_factuality} by Theorem 4.1 of \citet{mohri2024languagemodelsconformalfactuality}. The core differences between Conformal-RAG and \cite{mohri2024languagemodelsconformalfactuality} are the relevance function $R(c)$ used for filtering which incorporates similarity information from the RAG mechanism, and the use of automatic annotation to provide ground truth on sub-claim factuality.

\subsection{Conditional Conformal Factuality with RAG}
\label{subsec:conditional_converage_rag}
Previous research~\cite{foygelbarber2020limits, gibbs2024conformalpredictionconditionalguarantees} shows that marginal CP can undercover some groups within the data, while overcovering others, leading to fairness concerns \cite{romano2020with, cresswell2025conformal}. To address this, one can aim to provide group-conditional coverage over a pre-specified grouping $g: X\to G = \{1 \dots n_g\}$:
\begin{equation}
\label{eq:conditional}
    \mathbb{P}(y^*_\text{test} \in C_{\hat q_a}(x_\text{test}) \mid g(x_\text{test})=a)\geq 1-\alpha \quad \forall \ a\in G.
\end{equation}
Correspondingly, the conformal threshold $\hat q_a$ needs to depend on the group attribute $a$ (e.g. topic category or difficulty of the query). 

\citet{cherian2024llmenhancedconformal} proposed to adapt the threshold per test datapoint. First, define the pinball loss $\ell_{\alpha}(r) := (1 - \alpha)[r]_{+} + \alpha[r]_{-}$. Then, the threshold specific to datapoint $x_\text{test}$ is determined by the function $f_{\text{test}}: G\to \mathbb{R}$ defined as
\begin{equation}
f_{\text{test}}=\underset{f \in \mathcal{F}}{\arg\min} \frac{1}{n+1}\big[ \sum_{i=1}^n \ell_{\alpha}\big(S_i - f(g(x_{i}))\big) + \ell_{\alpha}\big(S_\text{test} - f(g(x_{\text{test}}))\big)\big]
\label{eq:quantile} 
\end{equation}
where $S_i = S(x_i, \hat{y}_{i})$ (\cref{eq:conformal_factuality_score}), $S_\text{test}$ is imputed using quantile regression, and the optimization is over the family of linear functions $\mathcal{F} = \{f(a) = \beta^\top e_{a}\}$ for $\beta\in \mathbb{R}^{|G|}$ and $e_a$ a basis vector of $\mathbb{R}^{|G|}$. The learned function $f_\text{test}$ provides the adapted conformal quantile $\hat q_{\text{test}}= f_\text{test}(x_\text{test})$ which is used to filter out claims, i.e. the method returns $y_\text{test}=F_{\hat q_{\text{test}}}(\hat y)$ as in \cref{eq:filter}. This procedure satisfies group-conditional factuality \cite{cherian2024llmenhancedconformal},
\begin{equation}
    \mathbb{P}(y^*_\text{test} \Rightarrow  y_\text{test}(x_\text{test}; \hat q)\mid g(x_\text{test})=a) \geq 1-\alpha\quad \forall \ a\in G.
    \label{eq:group_conformal_factuality}
\end{equation}
However, this method borders on impractical as it requires both a quantile regression to impute $S_\text{test}$, and an optimization over $\mathcal{F}$ for every inference datapoint. To simplify these procedures, Conformal-RAG follows the Mondrian CP paradigm \cite{vovk2003mondrian, vovk2005algorithmic} which first partitions the calibration data by groups using $g$, then calibrates a distinct threshold $\hat q_a$ for each $a\in G$ using the procedure in \cref{subsec:marginal_conformal_rag}. At inference time, the threshold for group $a_\text{test} = g(x_\text{test})$ is used for filtering out claims, i.e. we return $y_\text{test}=F_{\hat q_{a_\text{test}}}(\hat y)$. Since each group is calibrated independently, \cref{eq:conformal_factuality} holds for each group, which implies \cref{eq:group_conformal_factuality}.

\section{Experiments}
\label{sec:experiment}
\subsubsection*{Dataset}

We evaluate Conformal-RAG on four benchmark datasets: FActScore \cite{min2023factscorefinegrainedatomicevaluation}, PopQA \cite{mallen-etal-2023-trust}, HotpotQA \cite{yang2018hotpotqadatasetdiverseexplainable}, and MedLFQA \cite{jeong2024olaph}. The first three datasets use common knowledge from Wikipedia, whereas MedLFQA is a medical QA benchmark broken into five sub-datasets organized by topic and is considered more difficult than Wikipedia datasets for RAG. We follow the document curation process from previous work \cite{cherian2024llmenhancedconformal} for MedLFQA. For marginal Conformal-RAG, we evaluate the model on each of the four datasets individually. 
For conditional Conformal-RAG, we create a Wiki dataset by combining PopQA and HotpotQA, treating each as an individual group, while the MedLFQA dataset is divided into its underlying sub-datasets. In our experiment, the group labels are available during inference.

\subsubsection*{Experimental Setup}
For our experiments, we use a RAG system with a FAISS retriever~\cite{douze2025faisslibrary} and GPT-4o generator. For conformal calibration and inference, we adapted code from \citet{mohri2024languagemodelsconformalfactuality}. In addition, we use a GPT-4o model for annotation, sub-claim decomposition, and sub-claim merging as described in \cref{subsec:e2e_application}. For both marginal and conditional experiments, we test a range of error rates $\alpha\in [0.05, 0.40]$ and compare Conformal-RAG primarily to conformal factuality using confidence scoring directly from an LLM \cite{mohri2024languagemodelsconformalfactuality}. For clarity, we refer to our method as Conformal-RAG and the baseline as Conformal-LLM.

\subsection{Results}

\begin{figure*}
    \centering
    \includegraphics[width=\linewidth]{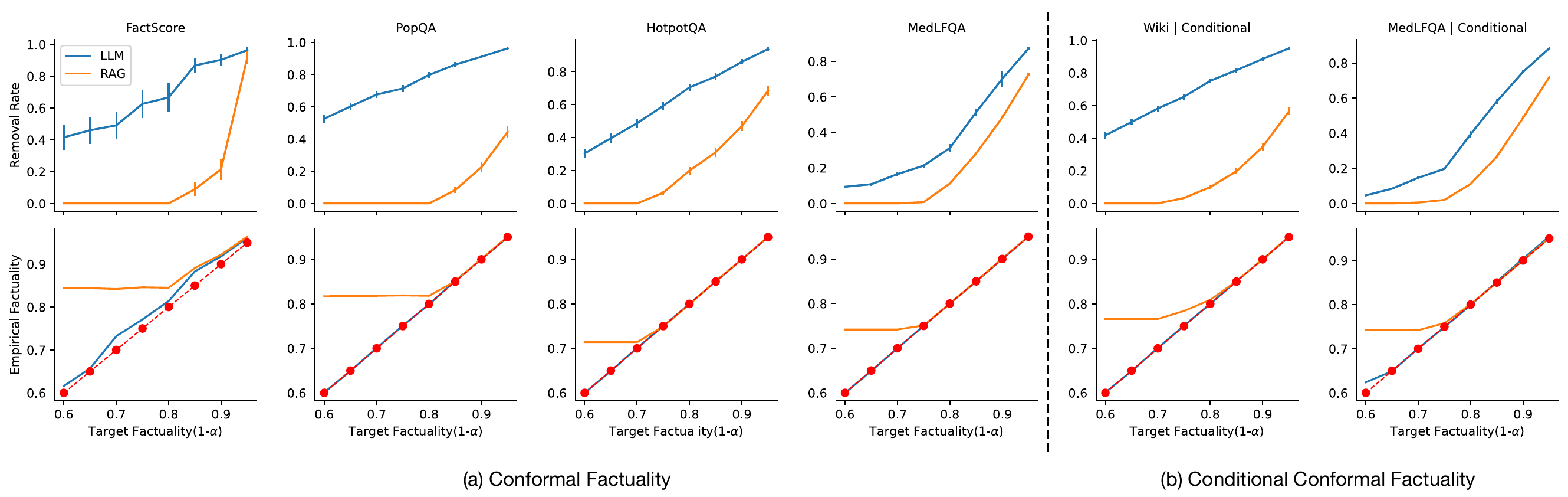}
    \vspace{-6mm}
    \caption{Sub-claim removal rates (top) and empirical factuality levels (bottom) for target factuality levels $1-\alpha$ using (a) marginal conformal prediction and (b) group-conditional conformal prediction, averaged over all test data. LLM is the baseline, while RAG is our method. The red dashed line shows the conformal factuality lower bound. }
    \label{fig:all_in_one}
\end{figure*}

\subsubsection*{Marginal Conformal Factuality}

We plot the removal rate and empirical factuality achieved with different target factuality levels $1-\alpha$ for both Conformal-RAG and Conformal-LLM in~\cref{fig:all_in_one}~(a). For removal rate, Conformal-RAG consistently outperforms the baseline, which only uses an LLM's parametric knowledge, across all four datasets. For example, Conformal-RAG's removal rate at target factuality level 85\% for FActScore is only 8.9\%, while Conformal-LLM removes 86.8\% of sub-claims to guarantee the same factuality level. Hence, Conformal-RAG is able to return longer, more informative answers with the same guarantees on factuality. For empirical factuality, calculated as the average factuality using the ground-truth labels from the test data, we find that both Conformal-RAG and Conformal-LLM maintain a level at or above the target, as expected from the guarantee in \cref{eq:conformal_factuality}. Hence, Conformal-RAG does not sacrifice factuality even when retaining a much higher fraction of claims. Notably, in many cases Conformal-RAG reaches a plateau of empirical factuality when the target $1-\alpha$ is lowered enough. In these cases, essentially all claims can be retained because the RAG mechanism does not generate as many non-factual claims in the first place. This clearly demonstrates the advantages of grounding generation in domain knowledge.

The design of our relevance scoring function from \cref{subsec:e2e_application} also benefits the quality of retained claims. At the individual data point level, we observe that Conformal-RAG preferentially filters out claims that may be factually correct, but lack semantic or contextual relevance to the given query. 
For example, on the query "how soon can tylenol be taken after a cocktail?" from MedLFQA (\cref{fig:demo}), one sub-claim states "there is no exact wait time specified [for alcohol metabolism]", which is factual but not relevant to the original question. This claim had low relevance $R(c) = 0.231$, leading to its removal at a relatively low target factuality of 75\%, corresponding to a threshold of $\hat q = 0.257$. By comparison, claims with higher relevance like "The UK's National Health Service suggests that a small amount of alcohol while taking acetaminophen is usually safe" with higher score $R(c) =0.472$, are both factual and more directly helpful for answering the query.

\subsubsection*{Conditional Conformal Factuality}
\begin{figure}
    \centering
    \includegraphics[width=0.45\linewidth]{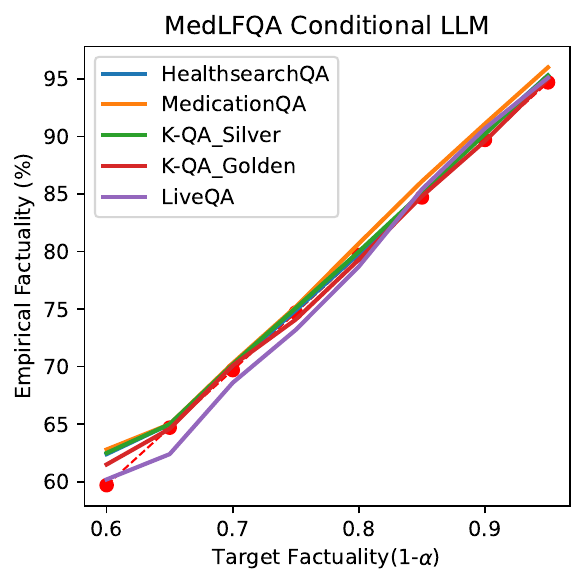} \includegraphics[width=0.45\linewidth]{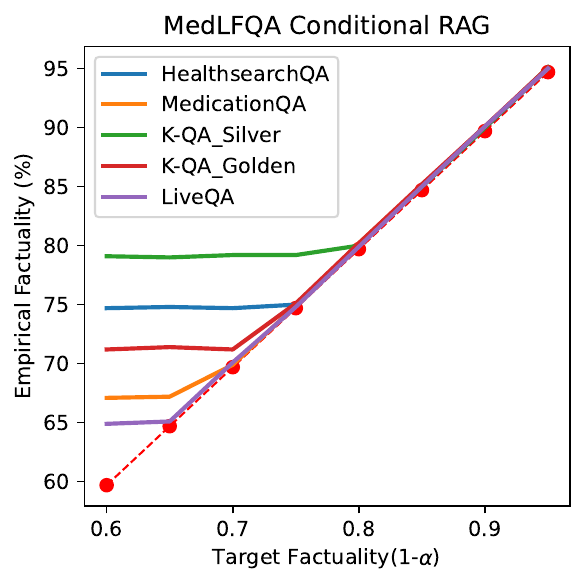}
    \caption{Empirical factuality by group for Conformal-LLM and Conformal-RAG on MedLFQA. The red dashed line shows the conformal factuality lower bound.}
    \label{fig:per-group-factuality-medLFQA}
\end{figure}

In~\cref{fig:all_in_one}~(b) we show results for conditional Conformal-RAG. We again observe that Conformal-RAG significantly reduces the removal rate while maintaining the (marginal) factuality guarantee. We further show the empirical factuality for each group on the MedLFQA dataset in~\cref{fig:per-group-factuality-medLFQA}. Both Conformal-LLM and Conformal-RAG approximately achieve the target factuality for every group, demonstrating their effectiveness across different subsets. However, Conformal-LLM shows slightly more variation, with some groups experiencing more deviation from the target factuality. For example, LiveQA shows a slight drop in factuality below the target when $1-\alpha<0.8$. In contrast, Conformal-RAG exhibits less fluctuation in factuality across the groups, suggesting more stable performance. This stability can likely be attributed to the effective use of RAG’s internal retrieval mechanism, which enhances the model’s consistency in achieving the target factuality. 

\section{Conclusion}
This paper introduced Conformal-RAG, a novel framework that applies conformal prediction (CP) to enhance RAG systems. An extension of Conformal-RAG to conditional CP  ensures group-conditional coverage across multiple sub-domains without requiring manual annotation of conformal sets, making it well-suited for complex RAG applications. Experimental results showed that Conformal-RAG and its conditional extension retain up to 60\% more high-quality sub-claims than direct applications of CP to LLMs, while maintaining the same factuality guarantees. 

\newpage
\bibliographystyle{ACM-Reference-Format}
\balance{}
\bibliography{ref}

\end{document}